\documentclass{PoS}

\title{Photon-jet correlations in hadronic collisions}

\ShortTitle{Photon-jet correlations in hadronic collisions}

\author{\speaker{Tomasz Pietrycki}\\
        Institute of Nuclear Physics, PL-31-342 Cracow, Poland\\
        E-mail: \email{Tomasz.Pietrycki@ifj.edu.pl}}

\author{Antoni Szczurek\\
        Institute of Nuclear Physics, PL-31-342 Cracow, Poland\\
        and University of Rzesz\'ow, PL-35-959 Rzesz\'ow, Poland\\
        E-mail: \email{Antoni.Szczurek@ifj.edu.pl}}

\abstract{
We compare results of $k_t$-factorization approach and next-to-leading order
collinear-factorization approach for photon-jet correlations in $pp$ and
$p \bar p$ collisions at RHIC, Tevatron and LHC energies. We discuss correlations 
in azimuthal angle 
as well as in two-dimensional space of transverse momenta
of photon and jet. Different unintegrated parton distributions (UPDF)
are included in the $k_t$-factorization approach. The results depend on
UPDFs used. The results of NLO collinear-factorization are shown for comparison.
}

\FullConference{High-pT physics at LHC\\
		 March 23-27 2007\\
		 University of Jyv\"askyl\"a, Jyv\"askyl\"a, Finland}

\begin{document}
%%%%%%%%%%%%%%%%%%%%%%
\section{Introduction}
%%%%%%%%%%%%%%%%%%%%%%
The jet-jet correlations are interesting probe of QCD dynamics
\cite{D0_dijets}. Recent studies of hadron-hadron correlations at
RHIC \cite{RHIC_hadron_hadron} open a new possibility to study the
dynamics of jet and particle production. The hadron-hadron correlations
involve both jet-jet correlations as well as complicated jet structure.
Recently also preliminary data on photon-hadron azimuthal correlation
in nuclear collisions were presented \cite{RHIC_photon_hadron}.
In principle, such correlations should be easier for theoretical
description as here only one jet enters, at least in leading order pQCD.
On the experimental side, such measurements are more difficult
due to much reduced statistics compared to the dijet studies.

Up to now no theoretical calculation for photon-jet were presented in
the literature, even for elementary collisions.
In leading-order collinear-factorization approach
the photon and the associated jet are produced back-to-back.
If transverse momenta of partons entering the hard process are
included, the transverse momenta of the photon and the jet are no
longer balanced and finite (non-zero) correlations in a broad range of
relative azimuthal angle and/or in lengths of transverse momenta of 
the photon and the jet are obtained. The finite correlations can be also
obtained in higher-order collinear-factorization approach \cite{Berends}.
According to our knowledge no detailed studies of photon-jet correlations
were presented in the literature.

%In contrast to the coincidence studies the inclusive distributions
%of photons were studied carefully up to next-to-leading order pQCD
%\cite{Aurenche87}.
%Similar studies were performed recently also in the $k_t$-factorization
%approach \cite{LZ_photon,PS06_photon}. A rather good description of
%direct-photon inclusive cross sections can be obtained in both approaches. 
%The $k_t$-factorization approach offers a relatively easy method to
%calculate the photon-jet correlations \cite{PS06_photon}
%The $k_t$-factorization approach 
%and was used recently to several
%high-energy reactions, including heavy quark pair photo-
%\cite{LS04,Mariotto} and hadroproduction \cite{BS00,LS06},
%charmonium production \cite{HKSST1,HKSST2}, inclusive $Z_0$
%\cite{KS04} and Higgs \cite{LZ05,LS05} production.

%===========================
\section{Formalism}
%===========================

It is known that at midrapidities and at relatively small transverse momenta
the photon-jet production is dominated by (sub)processes initiated by gluons.

In the $k_t$-factorization approach the cross section for a simultaneous
production of a photon and an associated jet in the collisions of two hadrons
($pp$ or $p \bar p$) can be written as
\begin{eqnarray} 
\frac{d\sigma_{h_1h_2 \to \gamma k}}{d^2p_{1,t}d^2p_{2,t}}
&=& \int dy_1dy_2 \frac{d^2k_{1,t}}{\pi}\frac{d^2k_{1,t}}{\pi}
\frac{1}{16\pi^2(x_1x_2s)^2}\overline{|{\mathcal M}_{ij\to \gamma k}|^2}\nonumber \\
&\cdot& \delta^2(\vec{k}_{1,t}+\vec{k}_{2,t}-\vec{p}_{1,t}-\vec{p}_{2,t})
{\cal F}_i(x_1,k_{1,t}^2,\mu_1^2) {\cal F}_j(x_2,k_{2,t}^2,\mu_2^2)  \; ,
\label{photon_jet_cross_section}
\end{eqnarray} 
where ${\cal F}_i(x_1,k_{1,t}^2,\mu_1^2)$ and ${\cal F}_j(x_2,k_{2,t}^2,\mu_2^2)$
are so-called unintegrated parton distributions.
The longitudinal momentum fractions are evaluated as $x_1=(m_{1t}
{\mathrm e}^{+y_1}+m_{2t}{\mathrm e}^{+y_2})/\sqrt{s}$ and
$x_2=(m_{1t}{\mathrm e}^{-y_1}+m_{2t}{\mathrm e}^{-y_2})/\sqrt{s}$. 
%
%We shall return to the choice of the factorization scale in the next section.
%Its role is completely different in different approaches
%i.e. different choices of UPDFs. A special attention will be devoted to
%the Kwieci\'nski UPDF and the role of the scale paramater. 
The final partonic state is $\gamma k = \gamma g, \gamma q$. 
If one makes the following replacement ${\cal F}_i(x_1,k_{1,t}^2)
\rightarrow x_1p_i(x_1)\delta (k_{1,t}^2)$, ${\cal F}_j(x_2,k_{2,t}^2) 
\rightarrow x_2p_j(x_2)\delta (k_{2,t}^2)$
and ${\mathcal M}_{ij \to \gamma k}(k_{1,t}^2,k_{2,t}^2) \to 
{\mathcal M}_{ij \to \gamma k}(k_{1,t}^2=0,k_{2,t}^2=0)$ 
then one recovers the standard leading-order collinear formula.

Up to now we have concentrated only on processes with two explicit 
hard partons ($\gamma k$) in the $k_t$-factorization approach.
It is of interest to compare results of our approach with the standard
collinear next-to-leading order approach.
%In this section we discus processes with three explicit hard partons. 

The cross section for $h_1 h_2\to \gamma k l X$  processes can be calculated
according to the standard parton model formula
\begin{eqnarray}
d\sigma_{h_1h_2\to \gamma kl} &=& \sum_{ijkl}
\int dx_1dx_2 \; p_i(x_1,\mu^2)p_j(x_2,\mu^2) \;
d\hat{\sigma}_{ij\to \gamma kl}    \; , 
\label{dsigma_h1h2}
\end{eqnarray}
where elementary cross section can be written as
\begin{eqnarray}
d\hat{\sigma}_{ij\to \gamma kl} = \frac{1}{2\hat s}
\overline{|{\mathcal M}_{ij\to \gamma kl}|^2} 
(2\pi)^4\delta^4(p_a + p_b - \sum_{i=1}^3 p_i) 
\prod_{i=1}^3\frac{d^3p_i}{2E_i(2\pi)^3} \; .
\label{dsigma_ij}
\end{eqnarray}
%
%This element can be expressed in an equivalent way in terms of parton 
%rapidities
%
%\begin{eqnarray}
%dR_3&=&(2\pi)^4\delta^4(p_a + p_b - \sum_{i=1}^3 p_i) 
%\prod_{i=1}^3\frac{dy_id^2p_{i,t}}{(4\pi)(2\pi)^2} \; ,
%\label{dR_3b}
%\end{eqnarray}
%
%The last formula is useful for practical applications. 

%Now the cross section
%for hadronic collisions can be written in terms of $2 \to 3$ matrix
%element as 
%
%\begin{eqnarray}
%d\sigma &=& \sum_{ijkl}
%dy_1d^2p_{1,t}dy_2d^2p_{2,t}dy_3 \frac{1}{(4\pi)^3(2\pi)^2}\frac{1}{\hat s^2}
%x_1p_i(x_1,\mu^2)x_2p_j(x_2,\mu^2)\overline{|{\mathcal M}_{ij \to \gamma
%  k l}|^2} \; ,
%\label{dsigma}
%\end{eqnarray}
%
%where the longitudinal momentum fractions are evaluated as
%$x_1= \frac{1}{\sqrt s} \sum_{i=1}^3 p_{i,t}{\mathrm e}^{+y_i}$,
%$x_2= \frac{1}{\sqrt s} \sum_{i=1}^3 p_{i,t}{\mathrm e}^{-y_i}$.
Repeating similar steps as for $2 \to 2$ processes we get finally
\begin{eqnarray}
d\hat{\sigma}_{ij \to \gamma kl}
&=&
\sum_{ijkl} \frac{1}{64\pi^4\hat s^2}x_1p_i(x_1,\mu^2)x_2p_j(x_2,\mu^2)
\overline{|{\mathcal M}_{ij \to \gamma k l}|^2} \\
&\times& 
p_{1,t}dp_{1,t}p_{2,t}dp_{2,t}d\phi_-dy_1dy_2dy_3 \; , 
\label{2to3_useful}
\end{eqnarray}
where the relative azimuthal angle between the photon and associated jet
$\phi_-$ is restricted to the interval $(0,\pi)$. The last formula 
is very useful in calculating the cross section for 
particle 1 and particle 2 correlations.

%======================
\section{Results}
%======================

In this section we shall present results for RHIC, Tevatron and LHC energies.
We use UPDFs from the literature.
There are only two complete sets of UPDFs in the literature
which include not only gluon distributions but also distributions
of quarks and antiquarks: (a) Kwieci\'nski \cite{Kwiecinski}, 
(b) Kimber-Martin-Ryskin \cite{KMR}.

For comparison we shall include also unintegrated distributions
obtained from collinear ones by the Gaussian smearing procedure.
Such a procedure is often used in the context of inclusive direct photon
production \cite{Owens,AM04}.
Comparing results obtained with those Gaussian distributions
and the results obtained with the Kwieci\'nski distributions with
nonperturbative Gaussian form factors will allow
to quantify the effect of UPDF evolution as contained in the
Kwieci\'nski evolution equations. What is the hard scale for our process?
In our case the best candidate for the scale 
is the photon and/or jet transverse momentum. Since we are interested
in rather small transverse momenta the evolution length is not
too large and the deviations from initial $k_t$-distributions
(assumed here to be Gaussian) should not be too big.

At high energies one enters into a small-x region, i.e. the region
of a specific dynamics of the QCD emissions. In this region only unintegrated
distributions of gluons exist in the literature. In our case
the dominant contributions come from QCD-Compton
$gluon-quark$ or $quark-gluon$ initiated hard subprocesses. 
This means that we need unintegrated distributions of both
gluons and quarks/antiquarks. In this case we take such UGDFs from the
literature and supplement them by the Gaussian distributions of 
quarks/antiquarks.

\begin{figure}[!htb] % Figure 1
\begin{center}
\includegraphics[height=6cm]{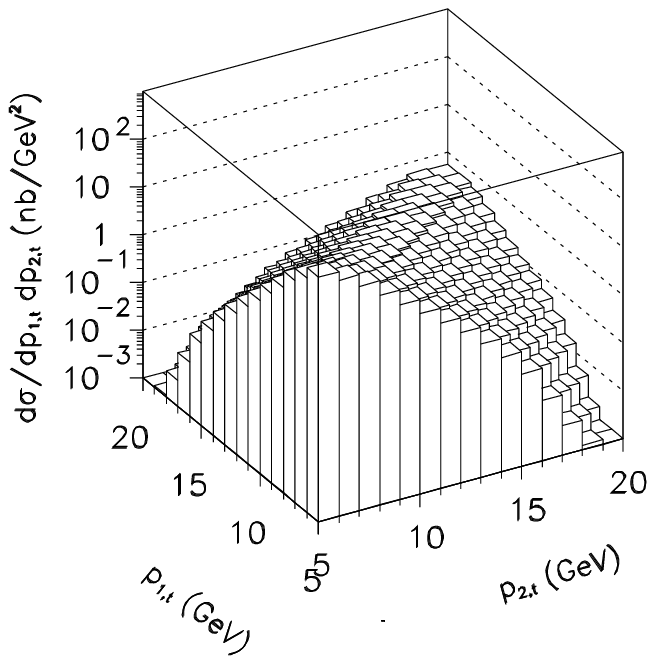}
\includegraphics[height=6cm]{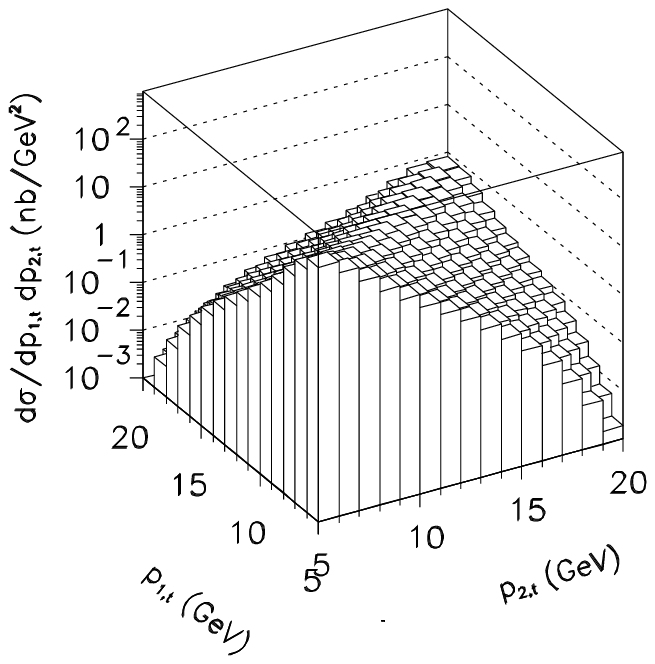}
\includegraphics[height=6cm]{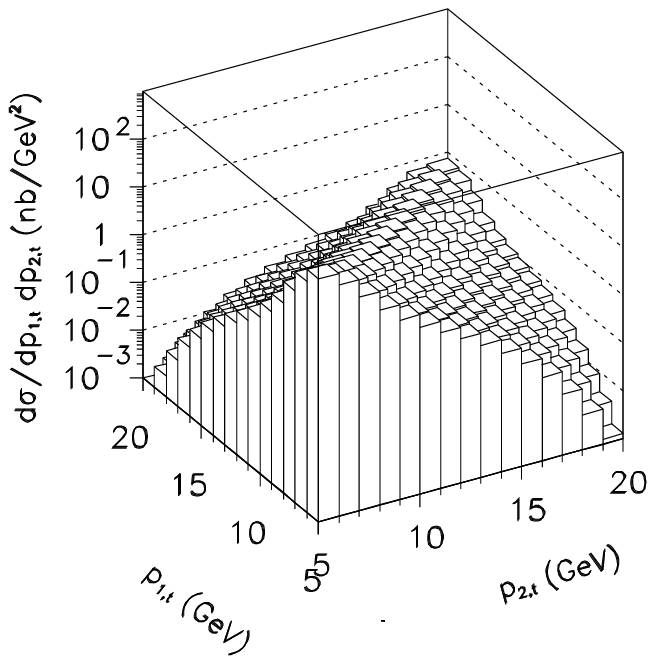}
\includegraphics[height=6cm]{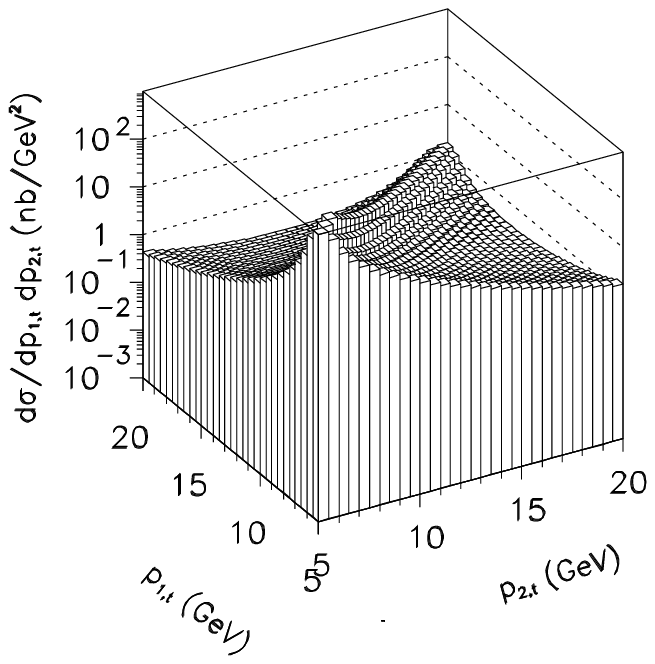}
\caption{
Transverse momentum distributions $d\sigma/dp_{1,t}dp_{2,t}$
at W = 1960 GeV and  
for different UPDFs in the $k_t$-factorization approach for
Kwieci\'nski ($b_0 = 1$~GeV$^{-1}$, $\mu^2 = 100$~GeV$^{2}$) (a), BFKL (b), KL
(c) and NLO $2 \to 3$ collinear-factorization approach including
diagrams from Fig.2. in Ref.\cite{PS07_correl} (d).
The integration over rapidities from the interval -5 $< y_1, y_2 <$
5 is performed.
\label{fig:updfs_ptpt}
}
\end{center}
\end{figure}

Let us start from presenting our results on the $(p_{1,t},p_{2,t})$ plane.
In Fig.\ref{fig:updfs_ptpt} we show the maps for different 
UPDFs used in the $k_t$-factorization approach as well as for NLO 
collinear-factorization approach for
$p_{1,t}, p_{2,t} \in (5,20)$~GeV and at the Tevatron energy $W =
1960$~GeV. In the case of the Kwieci\'nski distribution we have taken
$b_0$ = 1 GeV$^{-1}$ for the exponential nonperturbative form factor
and the scale parameter $\mu^2$ = 100 GeV$^2$.
Rather similar distributions are obtained for different UPDFs.
The distribution obtained in the collinear NLO approach differs qualitatively
from those obtained in the $k_t$-factorization approach.
First of all, one can see a sharp ridge along the diagonal $p_{1,t} = p_{2,t}$.
This ridge corresponds to a soft singularity when the unobserved
parton has a very small transverse momentum $p_{3,t}$.
As will be clear in a moment this corresponds to the azimuthal
angle between the photon and the jet being $\phi_{-} = \pi$. Obviously this is
a region which cannot be reliably calculated in collinear pQCD.
There are different practical possibilities to exclude this region from
the calculations.
The most primitive way (possible only in theoretical calculations) is to
impose a lower cut on transverse momentum of the unobserved parton $p_{3,t}$.
Secondly, the standard collinear NLO approach generates much bigger
cross section for asymmetric $p_{1,t}$ and $p_{2,t}$.
%We shall return to this observation in the course of this paper. 

%------------------------------------------------------------------
%------------------------------------------------------------------

\begin{figure}[!htb] % Figure 2
\begin{center}
\includegraphics[height=6cm]{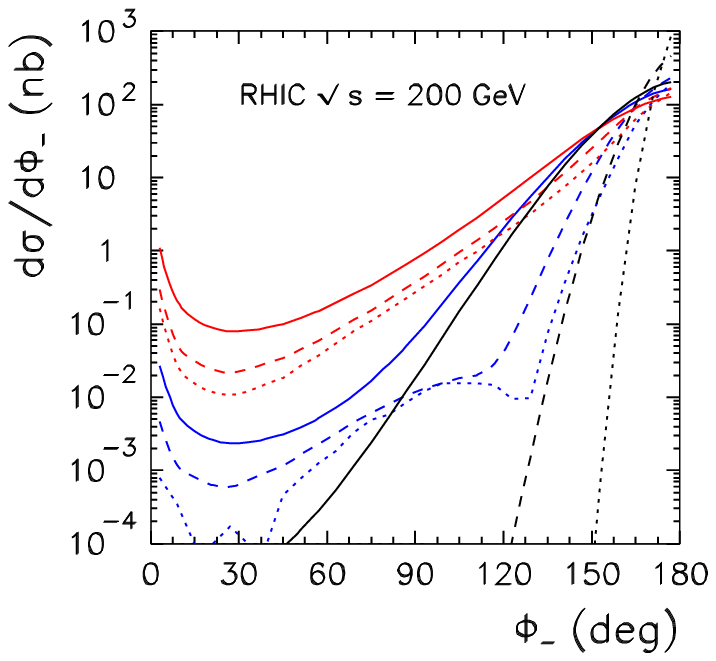}
\includegraphics[height=6cm]{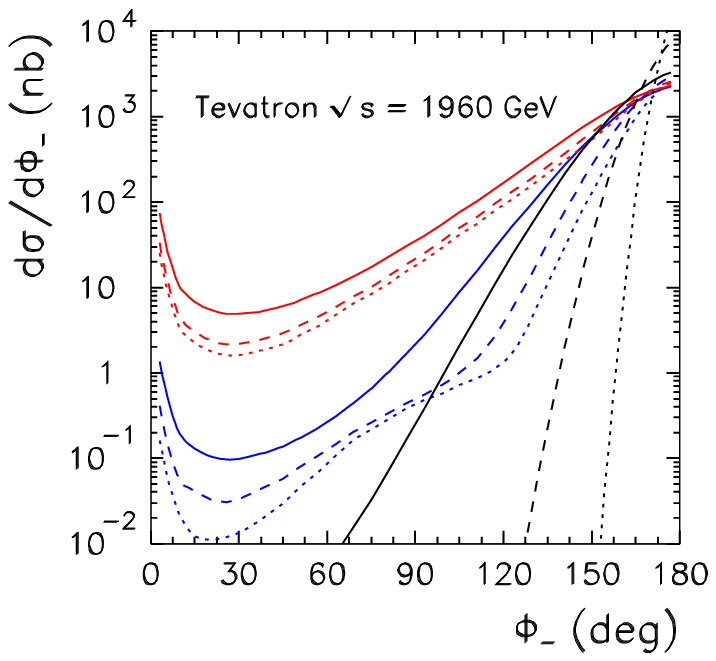}
\caption{
(Color on line) Azimuthal angle correlation functions at (a) RHIC, (b) Tevatron
energies for different scales and different values of $b_0$ 
of the Kwieci\'nski distributions.
The solid line is for $b_0 = 0.5$~GeV$^{-1}$, the dashed line is for
$b_0 = 1$~GeV$^{-1}$ and the dotted line is for $b_0 = 2$~GeV$^{-1}$.
Three different values of the scale parameters are shown: 
$\mu^2 = 0.25, 10, 100$~GeV$^2$ (the bigger the scale the bigger
the decorellation effect, different colors on line).
In this calculation  $p_{1,t}, p_{2,t} \in (5,20)$~GeV and
$y_1, y_2 \in$ (-5,5).
\label{fig:kwiecinski_scale}
}
\end{center}
\end{figure}

As discussed in Ref.\cite{PS06_photon} the Kwieci\'nski unintegrated parton
distributions are very useful to treat both the nonperturbative (intrinsic
nonperturbative transverse momenta)
and the perturbative (QCD broadening due to parton emission) effects on
the same footing.
In Fig.\ref{fig:kwiecinski_scale} we show the effect of
the scale evolution of the Kwieci\'nski UPDFs on the azimuthal angle
correlations between the photon and the associated jet.
We show results for different initial conditions ($b_0$ = 0.5, 1.0, 2.0
GeV$^{-1}$). At the initial scale (fixed here as in the original
GRV \cite{GRV98} to be $\mu^2$ = 0.25 GeV$^2$) there is a sizeable
difference of the results for different $b_0$. The difference
becomes less and less pronounced when the scale increases.
At $\mu^2$ = 100 GeV$^2$ the differences practically disappear.
This is due to the fact that the QCD-evolution broadening of
the initial parton transverse momentum distribution is much bigger than
the typical initial nonperturbative transverse momentum scale.

%------------------------------
\begin{figure}[!htb] % Figure 3 
\begin{center}
\includegraphics[height=4.6cm]{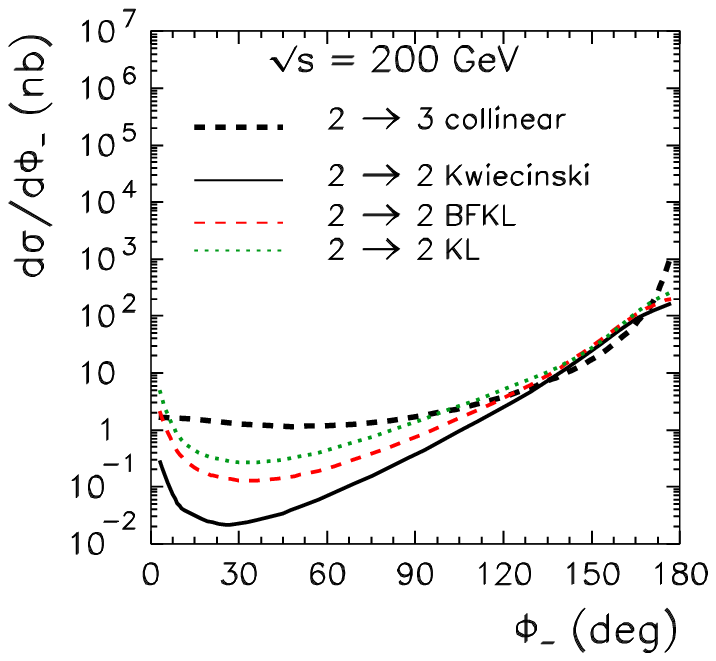}
\includegraphics[height=4.6cm]{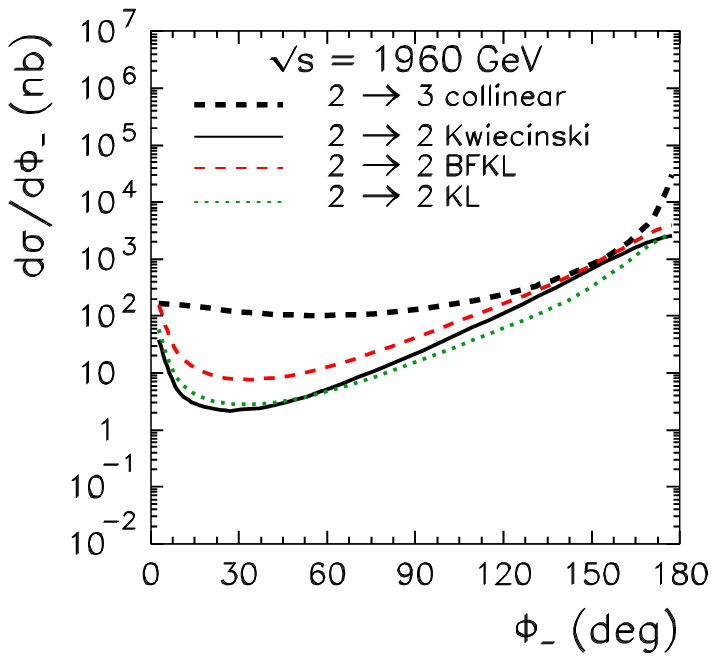}
\includegraphics[height=4.6cm]{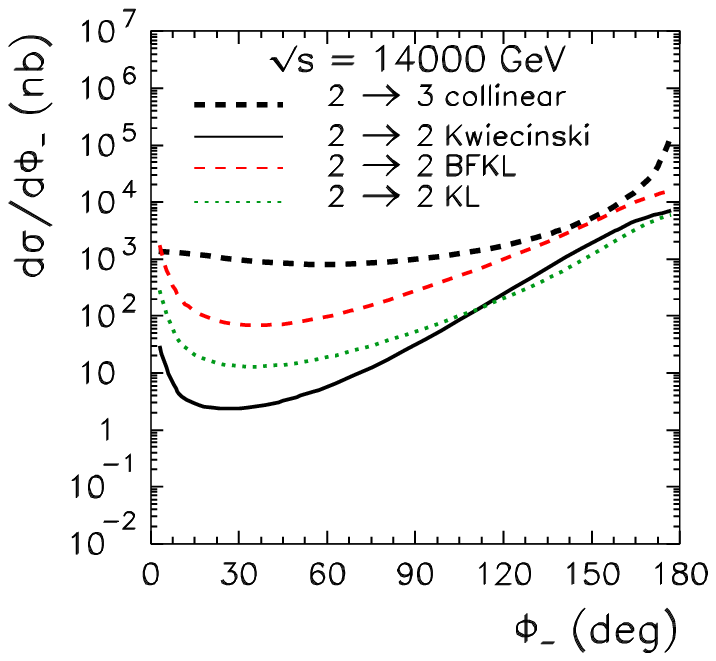}
\caption{
Photon-jet angular azimuthal correlations $d\sigma/d\phi_-$
for proton-(anti)proton collision at $W = 200, 1960, 14000$~GeV  
for different UPDFs in the $k_t$-factorization approach for
the Kwieci\'nski (solid), BFKL (dashed), KL (dotted) UPDFs/UGDFs
and for the NLO collinear-factorization approach (thick dashed).
Here $y_1, y_2 \in (-5,5)$.
\label{fig:updfs_phid}
}
\end{center}
\end{figure}

%-----------
In Fig.\ref{fig:updfs_phid} we show corresponding azimuthal angular
correlations for three different energies relevant for RHIC, Tevatron and LHC.
% again for different
%UPDFs in the $k_t$-factorization approach and NLO collinear-factorization
%approach.for the Tevatron energy $W = 1960$~GeV.
In this case integration is made over transverse momenta 
$p_{1,t}, p_{2,t} \in (5,20)$~GeV and rapidities $y_1, y_2 \in
(-5,5)$. The standard NLO collinear cross section grows somewhat faster with energy
than the $k_t$-result with unintegrated Kwieci\'nski distribution. This is
partially due to approximation made in calculation of the off-shell matrix elements.
Up to now we have used matrix elements called "on-shell" (for explanation see
appendix A in Ref.\cite{PS07_correl}). This approximation is expected to be 
reliable for small transverse momenta
of gluons (for a detailed discussion see Ref.\cite{PS06_photon}).
For larger gluon transverse momenta the longitudinal gluons start to play important
role. This is obiously not included in our simple extrapolation of the on-shell
formula.
%---------------------------------------------------------------------
%---------------------------------------------------------------------

\begin{figure}[!htb] % Figure 4
\begin{center}
\includegraphics[height=4.6cm]{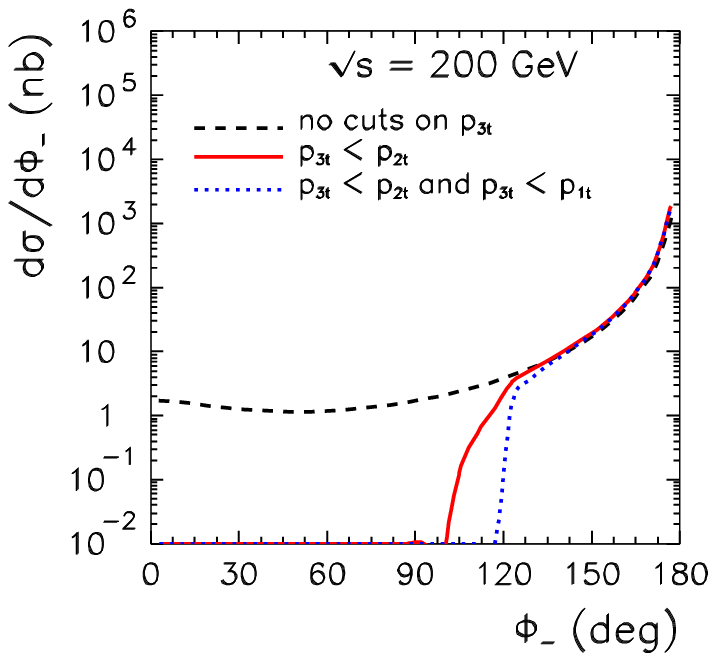}
\includegraphics[height=4.6cm]{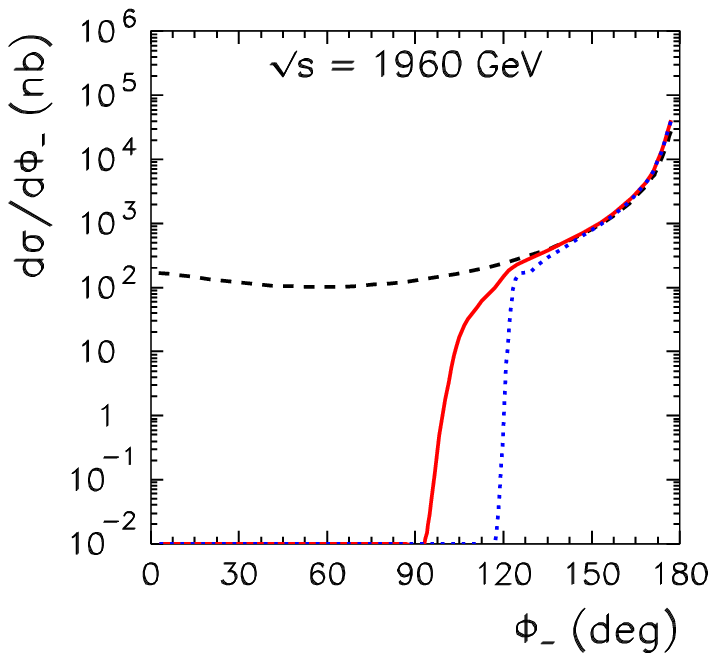}
\includegraphics[height=4.6cm]{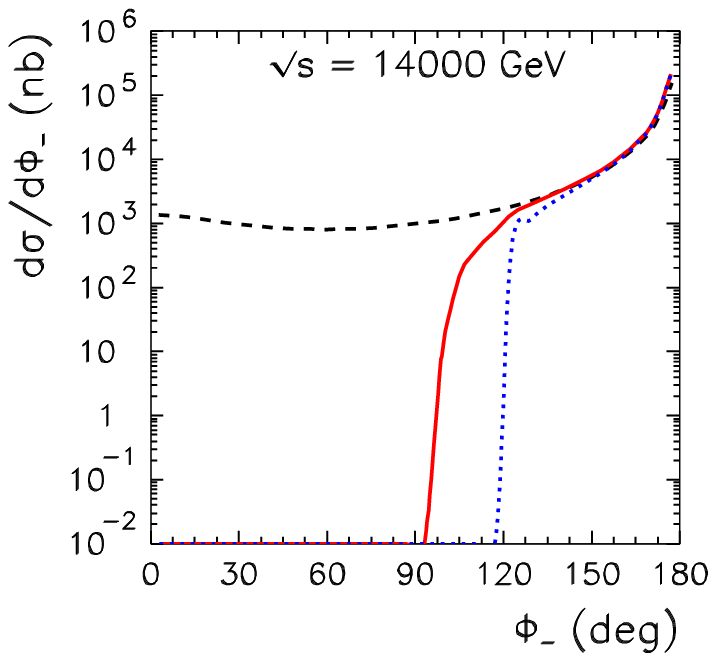}
\caption{
Angular azimuthal correlations in the NLO collinear-factorization approach
without any extra constraints (dashed), $p_{3,t} < p_{2,t}$ (solid),
$p_{3,t} < p_{2,t}$ and $p_{3,t} < p_{1,t}$ in addition (dotted).
Here $W = 200, 1960, 14000$~GeV and $y_1, y_2 \in (-5,5)$.
\label{fig:leading_jets_angle}
}
\end{center}
\end{figure}

In Fig.\ref{fig:leading_jets_angle} we show angular azimuthal
correlations for different relations between transverse momenta
of outgoing photon and partons:
(a) with no constraints on $p_{3,t}$, (b) the case where $p_{2,t} >
p_{3,t}$ condition (called leading jet condition in the following)
is imposed, (c) $p_{2,t} > p_{3,t}$ and
an additional condition $p_{1,t} > p_{3,t}$.
The results depend significantly on the scenario chosen as can be seen
from the figure. The general pattern is very much the same for different energies.

%------------------------------
%------------------------------

Fig.\ref{fig:leading_jets_angle} suggests that the $k_t$-factorization approach 
may be very efficient to describe correlation in azimuth for $\phi_- < \pi/2$
(where NLO contribution vanishes) when the leading jet condition is imposed.

%==============================
\section{Conclusions}
%==============================

We have performed for the first time the lacking in the literature
calculation of the photon-jet correlation observables in proton-(anti)proton
(RHIC, Tevatron and LHC) collisions.
Up to now such correlations have not been studied experimentally either.
We have concentrated on the
region of small transverse momenta (semi-hard region) where
the $k_t$-factorization approach seems to be the most efficient and
theoretically justified tool.
We have calculated correlation observables for different unintegrated parton
distributions from the literature. Our previous analysis of
inclusive spectra of direct photons suggests that the Kwieci\'nski
distributions give the best description of existing experimental data at low 
and intermediate energies.
%We have discussed the role of the evolution scale of the Kwieci\'nski
%UPDFs on the azimuthal correlations. In general, the bigger the scale
%the bigger decorrelation in azimuth is observed. When the scale
%$\mu^2 \sim p_t^2$(photon) $\sim p_t^2$(associated jet)
%(for the  kinematics chosen $\mu^2 \sim$ 100 GeV$^2$) is assumed, much bigger 
%decorrelations can be observed than from the standard Gaussian smearing
%prescription often used in phenomenological studies.

The correlation function depends strongly on whether it is the
correlation of the photon and any jet or the correlation of the photon 
and the leading jet.
In the last case there are regions in azimuth and/or in the two-dimensional
($p_{1,t}, p_{2,t}$) space which cannot be populated in the standard
next-to-leading order approach. In the latter case the $k_t$-factorization
seems to be a useful and efficient tool.

We hope that the photon-jet correlations will be measured
at Tevatron. At RHIC one can measure jet-hadron correlations
for rather not too high transverse momenta of the trigger photon and of
the associated hadron. This is precisely the semihard region discussed here.
In this case the theoretical calculations would require inclusion of the
fragmentation process. This can be done easily assuming independent parton
fragmentation method using fragmentation functions extracted
from $e^+ e^-$ collisions. This analysis is in progress.

\vskip 0.5cm

{\bf Acknowledgments}
We are indebted to Jan Rak from the PHENIX collaboration for the
discussion of recent results for photon-hadron correlations at RHIC.
This work was partially supported by the grant
of the Polish Ministry of Scientific Research and Information Technology
number 1 P03B 028 28.

%--------------------------

%---------------------
\end{document}